\documentclass[%
reprint,
superscriptaddress,
amsmath,amssymb,
aps,
prl,
]{revtex4-2}

\usepackage{graphicx}
\usepackage{dcolumn}
\usepackage{bm}
\usepackage[colorlinks=true, allcolors=blue]{hyperref}
\usepackage{parskip}
\usepackage{tikz,xcolor} 

\definecolor{lime}{HTML}{A6CE39}
\newcommand{\orcidicon}{%
	\begin{tikzpicture}
	\draw[lime, fill=lime] (0,0) 
	circle [radius=0.16] 
	node[white] {{\fontfamily{qag}\selectfont \tiny ID}};
	\draw[white, fill=white] (-0.0625,0.095) 
	circle [radius=0.007];
	\end{tikzpicture}
	\hspace{-2mm}
}

\newcommand{\orcid}[1]{\href{https://orcid.org/#1}{\orcidicon}}

\begin{document}
    \title{\texorpdfstring{High-field superconductivity from atomic-scale confinement \\ and spin-orbit coupling at (111)LaAlO$_3$/KTaO$_3$ interfaces}{High-field Superconductivity from Atomic-scale Confinement and Spin-orbit coupling at (111) LaAlO3/KTaO3 Interfaces}}

    \author{Ulderico Filippozzi\orcid{0000-0001-8146-0006}}
    \email{U.Filippozzi@tudelft.nl}
    \affiliation{Kavli Institute of Nanoscience, Delft University of Technology, Delft, The Netherlands}
    
    \author{Graham Kimbell}
    \affiliation{Department of Quantum Matter Physics, University of Geneva, Geneva, Switzerland}

	\author{Davide Pizzirani}
	\affiliation{High Field Magnet Laboratory (HFML - EMFL), Rabdoud university, Nijmegen, The Netherlands}
	
	\author{Siobhan McKeown Walker}
    \affiliation{Department of Quantum Matter Physics, University of Geneva, Geneva, Switzerland}
    \affiliation{Laboratory of Advanced Technology (LTA), University of Geneva, Geneva, Switzerland}

    \author{Chiara Cocchi}
    \affiliation{High Field Magnet Laboratory (HFML - EMFL), Rabdoud university, Nijmegen, The Netherlands}

    \author{Stefano Gariglio}
    \affiliation{Department of Quantum Matter Physics, University of Geneva, Geneva, Switzerland}

    \author{Marc Gabay}
    \affiliation{Laboratoire de Physique des Solides, Universite Paris Saclay, CNRS UMR 8502, Orsay Cedex, France}

    \author{Steffen Wiedmann}
    \affiliation{High Field Magnet Laboratory (HFML - EMFL), Rabdoud university, Nijmegen, The Netherlands}

    \author{Andrea D. Caviglia}
    \email{Andrea.Caviglia@unige.ch}
    \affiliation{Department of Quantum Matter Physics, University of Geneva, Geneva, Switzerland}

	\date{\today}

\begin{abstract}
    We study the superconducting critical fields of two-dimensional electron systems at (111)LaAlO$_3$/KTaO$_3$ interfaces as a function of electrostatic back-gating. Our work reveals in-plane critical fields of unprecedented magnitudes at oxide interfaces. By comparing the critical fields in-plane and out-of-plane we discover an extremely anisotropic superconductor with an effective thickness below 1~nm and a 12-fold violation of the Chandrasekhar-Clogston paramagnetic limit.
    The analysis of magneto-transport indicates that the enhancement of the critical fields is due to an exceptionally thin superconducting layer and to a paramagnetic susceptibility suppressed by spin-orbit scattering.
    
\end{abstract}
\maketitle
    Superconductivity -- the condensation of electrons into a zero-resistance macroscopic ground state -- only requires the breaking of the global gauge symmetry U(1). This minimal requirement results in a highly symmetric electron pairing in the form of a zero-momentum, spin-singlet state. However, in non-centrosymmetric crystals with strong spin-orbit coupling, the further reduced symmetry opens the door for unconventional pairing mechanisms, which can lead to enhanced critical fields or topological superconductivity \cite{Scheurer2015, Fukaya2018}. 
    In the quest to discover unconventional superconducting states, numerous material systems have been explored in both bulk and two-dimensional platforms \cite{Zhang_Falson_isingpairing, Bauer2012}. 
    Among these, KTaO$_3$-based superconductors have emerged as particularly promising candidates. Interfacial superconductivity in KTaO$_3$ \cite{Liu2021b, Chen2021, chen_electric_2021} shows a unique correlation between enhanced critical temperature and reduced symmetry at the interface, suggesting that inversion symmetry breaking is essential for the formation of the superconducting state \cite{Liu2023,chen2024,Mallik_2022,Gastiasoro2022a,Battacharya_tunability,Bruno2019}.
    As few other materials exhibit such a correlation, KTaO$_3$-based interfaces offer a unique platform to investigate the interplay between symmetry breaking, enhanced spin-orbit coupling, and their impact on the superconducting state. \\
    Here, we fabricate KTaO$_3$ two-dimensional electron gas (2DEG) devices and perform magnetotransport measurements, including measurements in static fields as high as 35~T. We observe extremely large critical magnetic fields at (111)LaAlO$_3$/KTaO$_3$ interfaces exceeding the Chandrasekhar-Clogston paramagnetic limit by a factor $\sim$12: amongst the largest  violations ever reported in two-dimensional superconductors \cite{Monteiro2017, Rout2017, Nakamura2019, Cao2021, Zhang_Falson_isingpairing, Ahadi_susc}. 
    In KTaO$_3$ interfaces and other 2D systems, such extremely large critical fields have been ascribed to an Ising-type superconducting pairing \cite{Zhang_Falson_isingpairing, Wickramaratne2020}, the emergence of an FFLO state \cite{Matsuda2007}, a vanishing spin susceptibility \cite{Ahadi_susc}, or unconventional spin-triplet superconducting pairing \cite{Zhang2013, Zhang2021a}. 
    Here we find that a combination of an extremely thin superconducting layer together with strong spin-orbit scattering are sufficient to suppress both orbital and paramagnetic pair-breaking mechanisms, leading to the observed critical fields. These results are central to the understanding of superconductivity in two-dimensional systems with strong spin-orbit coupling.

\begin{figure}[t]
        \centering
        \includegraphics[width=8.6cm]{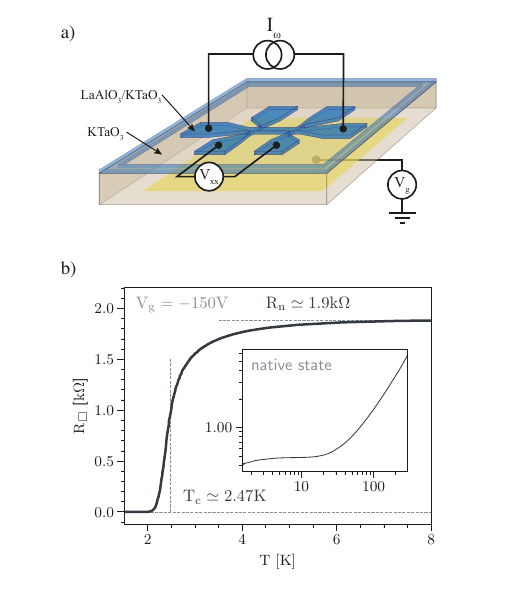}
        \caption{a) Hall bars are patterned in the electron gas to probe transport properties in two perpendicular in-plane directions $[1\bar{1}0]$ and $[11\bar{2}]$. A static back-gate voltage can be applied to the metallic backplate. b) A representative trace of sheet resistance vs temperature shows a clear superconducting transition with zero residual resistance. In the inset a trace of the resistance versus temperature for the sample's native state.}
        \label{fig:fig1}
    \end{figure}

    \textit{Results --} 
    Conducting samples were prepared via pulsed laser deposition (PLD) of amorphous LaAlO$_3$ (LAO) on (111)-oriented single-crystal KTaO$_3$ (KTO) substrates. Hall bar devices were fabricated using either pre-pattering or post-patterning techniques \cite{supp}. Owing to the large dielectric constant of the substrate ($\epsilon_r\simeq$ 4500 at low temperatures)\cite{Wemple1965}, the properties of the 2DEG can be effectively tuned through the application a static back-gate voltage between the top surface and a metallic back-plate to which the sample is thermally anchored.
    Fig.~\ref{fig:fig1}b shows the resistance versus temperature for a representative sample displaying a metallic behavior from room temperature down to $\simeq 5$~K, where the onset of a superconducting transition is revealed by enhanced conductivity due to superconducting fluctuations. 
    In the ``native state" (i.e. before any back-gate voltage is applied), the devices show an incomplete superconducting transition. However, after the ``gate-forming" process \cite{supp}, the critical temperature is raised and a clear transition is visible with very low residual resistance below 2~K (see Fig \ref{fig:fig1}b).
    
\begin{figure}[t]
        \centering
        \includegraphics[width=8.6cm]{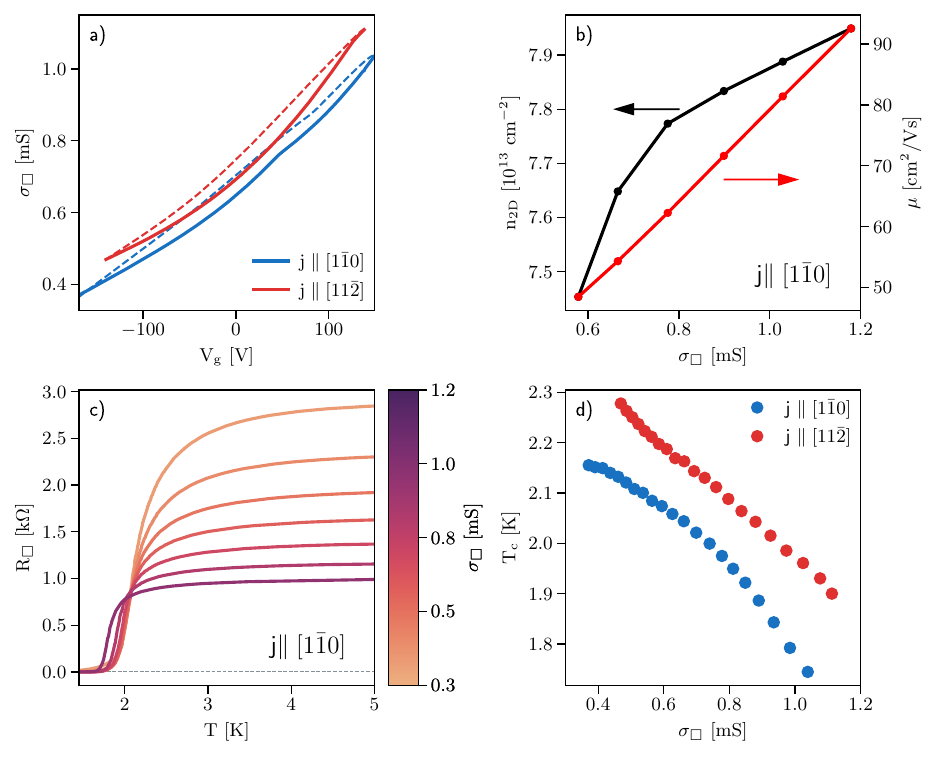}
        \caption{\textbf{Electric Field control of Superconductivity}. a) Sheet conductance as a function of electrostatic back-gate. The solid (dotted) lines represent the upward(downward) sweep branch. b) Two-dimensional carrier density and mobility measured for the different conductance states. c) Sheet resistance vs temperature traces at fixed back-gate voltage. d) Mean field critical temperature tuned by electrostatic back-gating, measured for both current directions.  }  
        
        \label{fig:fig2}
    \end{figure}

\begin{figure}[t]
        \centering
        \includegraphics[width=8.6cm]{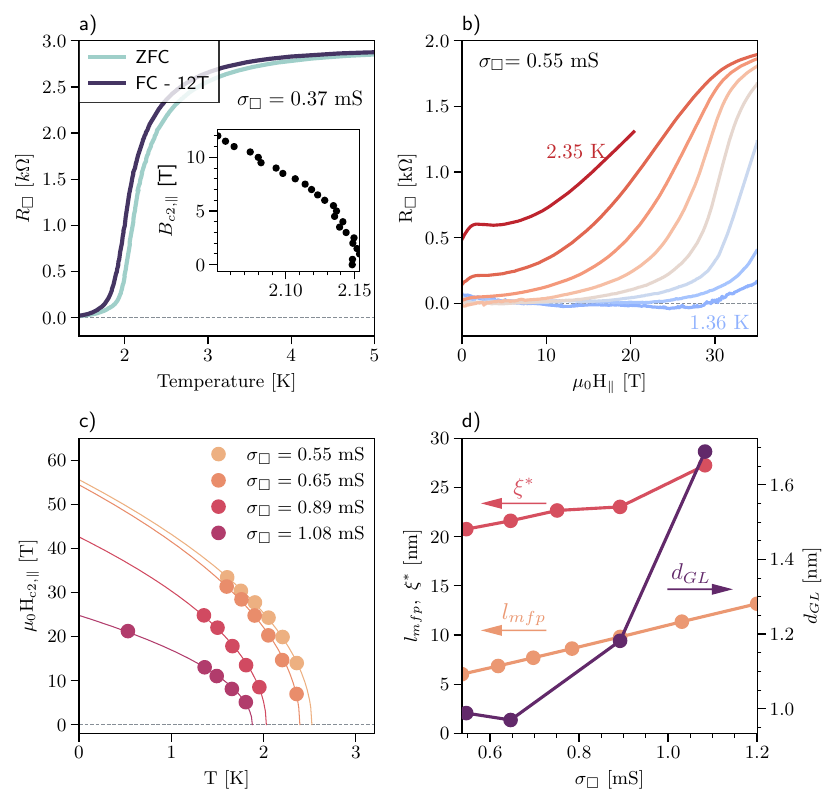}
        \caption{\textbf{Enhanced Critical Fields.} a) Comparison between Field-Cooldown (FC) and a Zero Field-Cooldown (ZFC) with applied in-plane magnetic field. Inset: the dependence of the in-plane critical field on temperature. b) Sheet resistance vs applied in-plane magnetic field for different temperatures. c) Measured in-plane critical field vs temperature for three different values of back-gate voltage. The solid line represents the 2D-Ginzburg-Landau model, the shaded regions correspond to the 1$\sigma$ confidence interval of the fit. d) Coherence length ($\xi^*$) and thickness of the superconducting layer ($d_\mathrm{GL}$) compared to the normal state mean free path ($l_\mathrm{MFP}$). }
        \label{fig:fig3}
    \end{figure}

\begin{figure*}[t]
            \centering
            \includegraphics[width = 15cm]{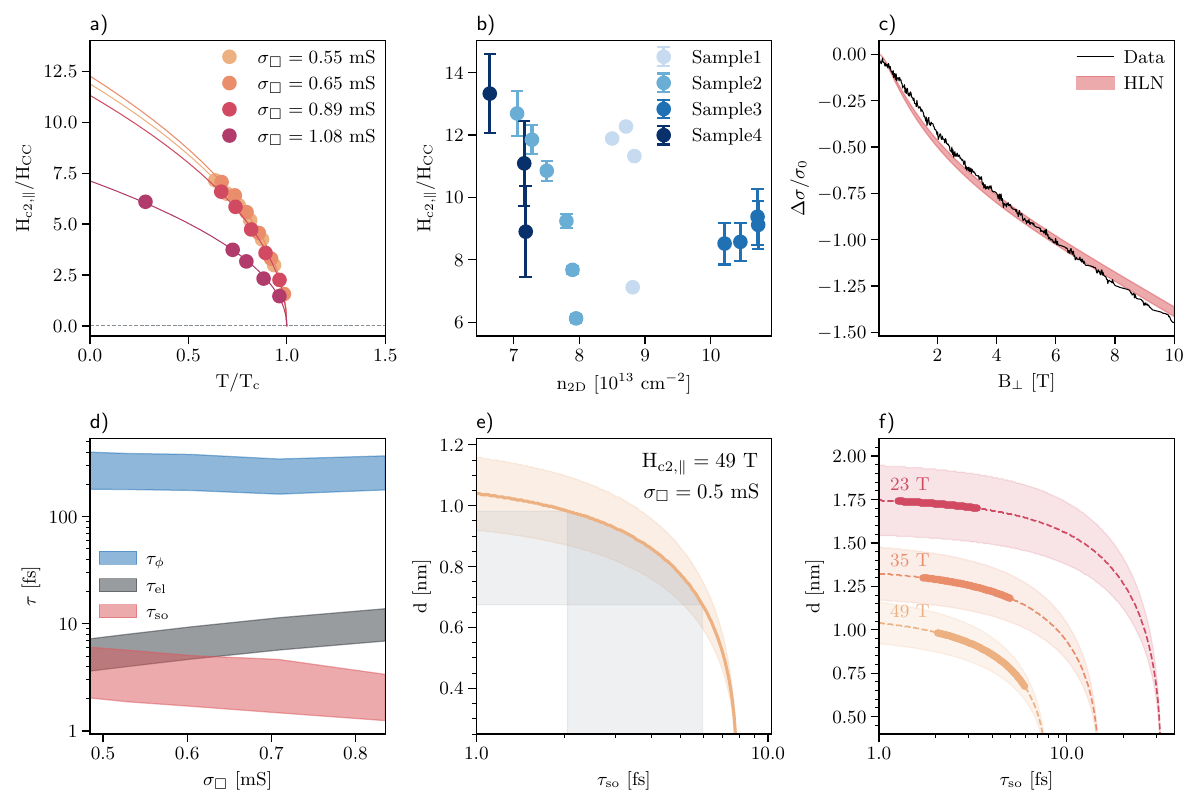}
            \caption{\textbf{Gate tunable violation of the Pauli Limit: a)} Temperature dependence of $H_\mathrm{c2,\parallel}$ normalized over the CC limit.\textbf{ b)} gate tuned violation of the CC limit for several samples as a function of the two dimensional carrier density. \textbf{c)} Magnetoconductance data overlaid with the HLN model computed in the range of optimal parameters \textbf{d)} Scattering times extracted from the magnetoconductance analysis. The color band represents the uncertainty in the determination of the optimum parameter as well as a range of effective electron masses $m^*\in [0.2, 0.4]m_\mathrm{e}$.  \textbf{e)}The yellow colorband represents the model from Eq.~\ref{eq2}. The uncertainty is dominated by the choice of effective mass within the orbital pair breaking term (solid line corresponds to $m^* = 0.3m_\mathrm{e}$). The gray shaded band represents the range of $\tau_\mathrm{so}$ and the corresponding range of $d$. \textbf{f)} The same analysis performed over a range of gate voltages. The dashed lines represent the model from equation \ref{eq2} while the dots highlight the range of $\tau_\mathrm{so}$ compatible with the magnetoconductance analysis.}
            \label{fig:fig4}
        \end{figure*}

    Figure ~\ref{fig:fig2}a shows an example of the back-gate tunability of the sheet conductivity. The change in sheet conductivity with back-gate voltage is due to a change in carrier concentration and mobility (Fig.~\ref{fig:fig2}b) which are determined from Hall effect measurements. 
    As previously reported \cite{chen_electric_2021}, the change in conductivity is mainly associated with a gate-induced variation of the Hall mobility ($\mu$) and only marginally with the change in the two-dimensional carrier density ($n_\mathrm{2D}$). Similar effects have been ascribed to back-gate induced modification of the confinement potential, which changes the effective disorder by pushing the electron gas closer to the more defective interface region \cite{chen_electric_2021,Hwang_dualgate,Bell_2009}.
    We measure the critical temperature $T_\mathrm{c}$ (defined here as $R(T_\mathrm{c})= R(10\mathrm{~K,\ 0~T})/2$) as it is reversibly tuned via electrostatic back-gating, see Fig.~\ref{fig:fig2}c,d. We observe that the critical temperature decreases with increased normal state conductivity, placing the samples in the over-doped region of KTO superconducting phase diagram \cite{chen_electric_2021, Battacharya_tunability}.
    The most striking feature of these devices is the robustness of superconductivity to the application of an in-plane magnetic field. Figure~\ref{fig:fig3}a compares a zero-field cooled (ZFC) to a field-cooled (FC) resistance vs temperature measurement with 12~T applied along the in-plane direction: $T_\mathrm{c}$
    is reduced by less than 100~mK (4\% $T_\mathrm{c}$). Figure~\ref{fig:fig3}b shows the magnetoresistance of the sample up to 35~T measured at different constant temperatures close to $T_\mathrm{c}$: the critical fields grow beyond the high magnetic field facility's capability already at $\approx T_\mathrm{c}/2$.
    We extrapolate the zero temperature critical field by fitting the data to the two-dimensional Ginsburg-Landau model (2D-GL): $H_{c2, \parallel} (T) = H_{c2, \parallel} \sqrt{1-T/T_c}$. As shown in Figure~\ref{fig:fig3}c, we find a good agreement between the 2D-GL model and the experiment up to fields of 35~T and down to temperatures of 450~mK. Notably, the extrapolated $H_{c2, \parallel}$ can reach values as high as 55~T, amongst the highest reported in interfacial oxide superconductor \cite{Ahadi_susc, Zhang2023a,Monteiro2017, Rout2017, Nakamura2019, Cao2021}.
    We measure the out-of-plane superconducting critical fields (See Supplementary Material \cite{supp}) and extrapolate their value at low temperature using the Ginzburg-Landau model ($H_{\mathrm{c2},\perp}(T) = H_{\mathrm{c2},\perp} (1-T/T_\mathrm{c})$).  We find $H_\mathrm{c2,\perp}$ varying as a function of gate approximately between 0.5 and 1~T, and extract the superconducting coherence length $\xi^*$ (see Figure ~\ref{fig:fig3}d). These reveal a coherence length $\approx4$ times longer than the mean-free-path ($l_\mathrm{MFP}$) confirming that the sample is in the dirty limit throughout the whole gating range.
    Finally, comparing $H_\mathrm{c2,\perp}$ to  $H_\mathrm{c2, \parallel}$ we compute the Ginzburg-Landau thickness of the superconducting layer $d_\mathrm{GL} = \sqrt{6\phi_0 H_\mathrm{c2, \perp}}/(\pi^{1/2} H_\mathrm{c2, \parallel})$ which falls below $1$~nm, corresponding to only few unit cells ($d_{(111)} = 0.23$~nm).
    All of these results were reproduced with different samples measured with our laboratory setup \cite{supp}.    
    \textit{Discussion --}We have presented critical fields of unprecedented magnitudes for interfacial oxide superconductors.
    In-plane critical fields of similar magnitude have been reported in systems such as ultra-thin van der Waals superconductors, with $T_\mathrm{c}$ spanning from several to tens of K. However, $T_\mathrm{c}$ hardly exceeds 2.5~K in our samples \cite{Zhang_Falson_isingpairing, Uchihashi_2017}, suggesting a robust protection against magnetic fields. \\
    To understand this effect, we must consider the possible Cooper pair-breaking mechanisms. The critical field of a superconductor is limited either by orbital or paramagnetic pair-breaking mechanisms.
    Orbital pair breaking is caused by the Lorentz force. With a strong enough magnetic field the kinetic energy gain of the electrons is sufficient to break Cooper pairs. The orbital limit can also be considered as a quantum of flux threading through the cross-sectional area of a Cooper pair ( $\propto \xi^{*2}$ or $\propto d\xi^*$ depending on the direction of the applied magnetic field). The in-plane orbital effect is suppressed in thin-films, as the number of out-of-plane momentum states is reduced, or the area of the Cooper pair is reduced.
    Paramagnetic pair breaking occurs because an applied magnetic field shifts the energy of spin-up and spin-down electrons. The limiting field occurs when the energy gain accompanied with superconducting pairing equals the energy gain of aligning the spins with the magnetic field, thus breaking the singlet Cooper pair. This trade off sets the so-called Chandrasekhar-Clogston (CC) limit\cite{Chandrasekhar_1962, Clogston_1962}: the maximum magnetic field ($\mu_0 H_\mathrm{CC}$) that a superconductor can tolerate. Assuming weak BCS-like-coupling, the CC field reads:
\begin{equation}
\label{eq1}
\mu_0 H_\mathrm{CC} = \frac{1.76 \ k_\mathrm{B} T_\mathrm{c}}{\mu_\mathrm{B} \sqrt{\ g_\mathrm{eff}}}  = 1.38 \bigg[\mathrm{\frac{T}{K}}\bigg] \times T_\mathrm{c}
\end{equation}
    where $g_\mathrm{eff}=2$ is the Land\'e $g$-factor. Our measurements reveal a gate tuneable violation of the CC limit which, in optimal configuration, can be more than tenfold (see Figure~\ref{fig:fig4}a).\\
    In the context of high field superconductivity in thin films, it is understood that the presence of spin-orbit coupling (SO) (elastic scattering that affects the spin channel) effectively reduces the paramagnetic susceptibility, suppressing the energy gain for aligning spins with the applied field and increasing the paramagnetic critical field \cite{WHH_1966, Maki1966, Fulde1973}. The interplay between confinement and SO-suppressed paramagnetic pair breaking was observed in thin films of aluminum with heavy-metals overlayers \cite{Tedrow1982} and in $\delta$-doped SrTiO$_3$ \cite{Kim2012}.
    Orbital and paramagnetic pair breaking give two parallel pathways for the suppression of superconductivity. By considering only the orbital component, we can find an upper bound for the effective thickness of the superconductor (Figure \ref{fig:fig3}d). A refinement of that estimate requires an independent measurement of the strength of spin-orbit scattering in these samples.\\
    We performed an analysis of weak anti-localisation (Figure~\ref{fig:fig4}c) using the Hikami-Larkin-Nagaoka (HLN) model \cite{Hikami1980} in order to independently quantify the spin-orbit scattering time $\tau_\mathrm{so}$.
    We find a good agreement between data and the HLN model. Due to uncertainties in the fitting of the HLN model, we can only provide a range of optimal parameters (see Figure~\ref{fig:fig4}d and \cite{supp}) which result in a good agreement between the model and data (see Figure~\ref{fig:fig4}c). The range provided for the scattering times accounts both for the uncertainty in the analysis and for a choice of effective masses ($m^* \in [0.2, 0.4]m_e$) that includes the varying values reported from spectroscopic measurements for the light electron band \cite{Bareille2014, mallik_electronic_2023, zhang_strain-driven_2018, Santander-Syro2012, Varotto2022}. 
    The analysis reveals an extremely short $\tau_\mathrm{so}$ that can even be comparable to the elastic scattering time $\tau_\mathrm{el}$ \cite{Caviglia2010}. Moreover, we observe $\tau_\mathrm{so}$ becoming shorter as the sample becomes more conducting. This trend of SO was already observed both in (001) and (111) interfaces of SrTiO$_3$ \cite{Caviglia2010, Rout2017} and later understood as a consequence of an enhanced Rashba splitting of the bands near avoided level crossings in the bands \cite{Joshua2012, Liang2015}.\\
    Figure \ref{fig:fig3}d also highlights another unique feature of the superconducting state in these samples: $d_\mathrm{GL}\ll l_\mathrm{MFP}$ placing these samples in the ultra-thin limit. In this regime the standard diffusive model for a dirty 2D superconductor is not applicable, and the orbital-limited critical field follows $H_{c2,\parallel}\propto d^{-3/2}$ \cite{supp, Parks1969, Thompson1965}.
    In order to treat orbital pair breaking in the ultra-thin limit as well as SO-suppressed paramagnetic pair breaking, we adopt the following model \cite{supp, Maki1966}   :
    \begin{equation}
    \label{eq2}
    H_{\mathrm{c},\parallel}(T) = \frac{2}{\sqrt{\pi}}\bigg(\frac{v_{\mathrm{F}}e^2d^3}{16\hbar k_{\mathrm{B}}T_{\mathrm{c}}} + \frac{3\tau_{\mathrm{so}}\mu_{\mathrm{B}}^2}{\hbar k_{\mathrm{B}}T_c}\bigg)^{-1/2}\bigg(1-\frac{T}{T_c}\bigg)^{1/2}
    \end{equation} 
    This model allows us to capture the interplay between orbital-limited critical fields (first term on the right hand side) and paramagnetic-limited critical fields (second term) with a SO-induced renormalization of the electronic susceptibility. We highlight that Eq.~\ref{eq2} clearly shows how the orbital and paramagnetic pair breaking are coupled rather than giving independent limiting fields as is sometimes assumed.
    In order to compare the model with the data, we first compute the set of values $(d, \tau_\mathrm{so})$ that yield the same $H_\mathrm{c2, \parallel}(0)$ given $v_\mathrm{F} = \hbar\sqrt{2\pi n_\mathrm{2D}}/m^*$ and $T_\mathrm{c}$ (which are measured independently).
    The results of this procedure is plotted in figure \ref{fig:fig4}e (solid line corresponding to $m^* = 0.3m_\mathrm{e}$). 
    For very short $\tau_\mathrm{so}$ the purely paramagnetic-limited critical field lies far above $H_\mathrm{c2, \parallel}(0)$, therefore orbital effects are dominating (i.e. $d$ is weakly dependent on $\tau_\mathrm{so}$). 
    On the other hand, as $\tau_\mathrm{so}$ increases and the paramagnetic-limited critical field decreases, both mechanisms become relevant. As two parallel pathways for Cooper pair breaking are available now, $d$ has to decrease accordingly to maintain the value of $H_\mathrm{c2,\parallel}(0)$. The SO scattering times extracted from the analysis of the weak anti-localisation (gray band) are overlaid with the model and result in a corresponding interval for $d$.\\
    The gate configuration that we chose here (Figure \ref{fig:fig4}e) nicely illustrates the interplay between orbital and paramagnetic effects: the estimated thickness $d$ depends strongly on the value of $\tau_\mathrm{so}$ showcasing the crucial contribution of paramagnetic pair breaking.
    As the analysis is extended to other gate configurations, we can see a trend from the highest critical field (more resistive state), where both pair breaking mechanisms are relevant, to the lowest critical fields (more conducting states) which are mainly limited by two-dimensional confinement through orbital effects. For the more conducting states, the increased SO scattering rate and reduced $H_\mathrm{c2,\parallel}(0)$ causes the estimation of $d$ to be almost unaffected by the choice of $\tau_\mathrm{so}$.\\
    In light of this discussion, we believe that the most likely scenario is that the parallel critical fields in these samples are limited by a combination of orbital and paramagnetic effects: spin-orbit coupling and two-dimensional confinement cooperate to bring the critical fields to magnitudes that are unprecedented for oxide interfaces.
    By comparing several samples, we do not find a trend compatible with \cite{Ahadi_susc}, i.e. a 9-fold violation of the CC Limit can be achieved at any carrier density between $6.6\times10^{13}$~cm$^{-2}$ and  $10.8 \times 10^{13}$~cm$^{-2}$ (see Figure~\ref{fig:fig4}b).\\
    We conducted further experiments to clarify the origin of these large critical fields. As K is relatively volatile, we checked if K-deficiency could be suppressing superconductivity. We pre-annealed a substrate in vacuum at high temperature before deposition, resulting in $\sim$30\% K-deficiency at the KTO surface, but the sample still showed a parallel critical field around 50~T (see Supplementary Material)\cite{supp}.
    Additionally, we do not find a measurable modulation of in-plane critical field with in-plane field angle (see Supplementary Material)\cite{supp}.
    We note that critical fields of this magnitude require careful parallel alignment of the sample to be observed and are not associated to particular conditions during sample preparation \cite{supp}. 
    Considering the estimated thickness of the superconducting layer it is possible that very unusually wide transition (Fig.~\ref{fig:fig3}b) could be caused by inhomogeneity of the SC layer's thickness due to the roughness of the substrate. The role of in-plane inhomogeneities remains to be investigated.

    \textit{Conclusion -- }We have fabricated superconducting two-dimensional electron systems at the surface of KTaO$_3$(111) and we studied their superconducting phase diagram via electrostatic back-gating.
    We observe critical fields of unprecedented magnitude in interfacial oxide superconductors which can reach up to 56~T, violating the Chandrasekhar-Clogston paramagnetic limit by more than ten times.
    Combining the weak antilocalisation and parallel critical fields analyses, we find both an extremely thin superconducting layer, suppressing orbital pair breaking, and extremely short spin-orbit scattering times, suppressing paramagnetic pair breaking. The combination of these two effects gives rise to the large critical fields observed in experiments.

\textit{Acknowledgments --}
We thank J.-M. Triscone, M. Cuoco, R. Citro and C. Ortix for fruitful discussions.
We acknowledge funding from the TOPCORE project (with project number OCENW.GROOT.2019.048) of the research program Open Competition ENW Groot financed by the Dutch Research Council (NWO). This work was supported by HFML-RU/NWO-I, a member of the European Magnetic Field Laboratory (EMFL), by the Swiss State Secretariat for Education, Research and Innovation (SERI) under contract no. MB22.00071, by the Gordon and Betty Moore
Foundation (grant no. 332 GBMF10451 to A.D.C.) and by the European Research Council (ERC).

\textit{Author Contribution --} 
U.F. and G.K. fabricated and characterized the samples. 
U.F., G.K., D.P. and C.C. performed the electrical transport measurements with support from S.G. and S.W.. 
S.M.W. performed the XPS measurements and analysis. 
U.F. performed the data analysis and produced the figures with support from G.K., S.M.W. and M.G.. 
M.G. provided theoretical support for the analysis of the magnetoconductance and critical fields.
A.D.C. supervised the project. 
All authors wrote the manuscript with a first draft provided by U.F., G.K. and S.M.W..

\textit{Data Availability -- }The raw data and scripts adopted for data analysis and the realization of the figures are available in the \href{https://doi.org/10.5281/zenodo.14055556}{Zenodo Repository}.

\bibliographystyle{unsrt}
\bibliography{references.bib}

\end{document}